\documentclass[a4paper,11pt]{article}
\usepackage{amsmath}
\usepackage{amsfonts}
\usepackage{amssymb}
\usepackage{bm}
\usepackage[utf8x]{inputenc}
\usepackage{ae}
\usepackage[T1]{fontenc}
\usepackage[english]{babel}
\usepackage{graphicx}
\usepackage{makecell}
\usepackage{placeins}
\usepackage{tabularx}
\usepackage{tabulary}
\usepackage{setspace}
\usepackage{tablefootnote}
\usepackage[flushmargin]{footmisc}
\usepackage[comma]{natbib}
\usepackage{array}
\usepackage{float}
\usepackage{caption}
\usepackage{tikz}
\usetikzlibrary{shapes.geometric}
\usepackage[a4paper,left=2.5cm,right=2.5cm,top=3cm,bottom=3cm]{geometry}
\usepackage[colorinlistoftodos]{todonotes}
\usepackage{pdfpages}
\usepackage{longtable}
\usepackage{adjustbox}
\usepackage{booktabs}
\usepackage{multirow}
\usepackage{lscape}
\usepackage[colorlinks=true, allcolors=blue]{hyperref}

\usepackage[flushleft]{threeparttable}
\usepackage{inputenc}



\begin{document} \doublespacing \pagestyle{plain}
	
	\def\ci{\perp\!\!\!\perp}
	\begin{center}
		
		{\LARGE Austria's KlimaTicket: Assessing the short-term impact of a cheap nationwide travel pass on demand
		}

		{\large \vspace{0.8cm}}
		
		{\large Hannes Wallimann}\medskip

		{\small {University of Applied Sciences and Arts Lucerne, Competence Center for Mobility} \bigskip }
		
		{\large \vspace{0.8cm}}
		
		{\large Version February 2024}\medskip

	\end{center}
	
	\smallskip

	\noindent \textbf{Abstract:} {Measures to reduce transport-related greenhouse gas emissions are of great importance to policy-makers. A recent example is the nationwide KlimaTicket in Austria, a country with a relatively high share of transport-related emissions. The cheap yearly season ticket introduced in October 2021 allows unlimited access to Austria's public transport network. Using the synthetic control and synthetic difference-in-differences methods, I assess the causal effect of this policy on public transport demand by constructing a data-driven counterfactual out of European railway companies to mimic the number of passengers of the Austrian Federal Railways without the KlimaTicket. The results indicate public transport demand grew slightly faster in Austria, i.e., 3.3 or 6.8 percentage points, depending on the method, than it would have in the absence of the KlimaTicket. However, the growth effect after the COVID-19 pandemic appears only statistically significant when applying the synthetic control method, and the positive effect on public transport demand growth disappears in 2022.

	}
	
	{\small \smallskip }
	{\small \smallskip }
	{\small \smallskip }
	
	
	{\small \smallskip }
	{\small \smallskip }
	{\small \smallskip }
	
	{\small \noindent \textbf{Acknowledgments:} Michael Stöckli provided excellent research assistance. I would like to thank Kevin Blättler, Silvio Sticher, and Widar von Arx for interesting discussions and helpful comments.}
	
	\bigskip
	\bigskip
	\bigskip
	\bigskip
	
	{\small {\scriptsize 
\begin{spacing}{1.5}\noindent  
\textbf{Addresses for correspondence:} Hannes Wallimann, University of Applied Sciences and Arts Lucerne, Rösslimatte 48, 6002 Lucerne, \href{mailto:hannes.wallimann@hslu.ch}{hannes.wallimann@hslu.ch}.
\end{spacing}
			
		}\thispagestyle{empty}\pagebreak  }

	{\small \renewcommand{\thefootnote}{\arabic{footnote}} %
		\setcounter{footnote}{0}  \pagebreak \setcounter{footnote}{0} \pagebreak %
		\setcounter{page}{1} }
	
\section{Introduction}\label{introduction}

The transport sector causes a large amount of all CO$_2$ emissions and energy consumption worldwide. With cycling and walking, public transportation is a sustainable alternative to private care use \citep{redman2013quality}. A simple way to increase the attractiveness of public transportation is to lower public transport fares through subsidization. The probably most famous example in Europe was the so-called 9-Euro-Ticket, which was introduced in the light of increasing energy costs in Germany \citep[see, e.g., ][]{aydin2023public,loder2024observing}. The 9-Euro-Ticket allowed passengers to use regional trains and urban public transportation infrastructures (buses, trams, and subway) throughout Germany in June, July, and August 2022 for nine Euros per month. Another example is the introduction of lower fares in Geneva, an urban area in Switzerland. There, electorates of the canton of Geneva chose to reduce the prices of state public transport in 2013 by, on average, 12.6\% \citep[see, e.g., ][]{wallimann2023price}. In the transportation literature, there exist many studies discussing the effect of fare policies on public transport demand \citep[see, among others, e.g., ][]{brechan2017effect,holmgren2007meta,shin2021exploring,kholodov2021public}.

When it comes to natural experiment estimates of price policies, the experiments often comprise regional or time-limited events such as the fare reduction in Geneva or the 9-Euro-Ticket. In contrast, the introduction of the so-called "KlimaTicket" in Austria is a nationwide and long-term policy. The KlimaTicket, introduced on October 26, 2021, allows individuals free and unlimited access to the public transport network in Austria at a fixed price of 1095 Euros for the classic category and 821 Euros for youths, seniors, and individuals with disabilities \citep[see, e.g., ][]{OECD_Klimaticket_2022}. In August 2023, 250,000 customers in Austria had already purchased a KlimaTicket, about 3\% of the adult population in Austria \citep{KlimaTicketReport2022}. The number increases to 15\% when also considering regional variants of the KlimaTicket in transport associations and federal states \citep{KlimaTicketReport2022}. The policy is of particular relevance because transport-related greenhouse gas emissions represented 39.8\% of total CO$_2$ emissions in Austria, which is noticeably higher than the share in the OECD area with 24\% \citep{OECD_Klimaticket_2022}.

While the effects of the KlimaTicket have already been discussed in first studies \citep[see, e.g., ][]{KlimaTicketReport2022,OECD_Klimaticket_2022}, I am the first to estimate the causal effect of the ticket on public transport demand. In this study, I show initial causal demand effects and how future analyses by researchers and practitioners can be structured to analyze the KlimaTicket or public transport price policies in general. In a comparative case study, I use the synthetic control and difference-in-differences methods, introduced by \citet{abadie2010synthetic} or \citet{arkhangelsky2019synthetic}, to construct a synthetic Austrian Federal Railways (ÖBB). This synthetic counterfactual---a weighted subset of comparable European public transport companies---mimics the demand the actual ÖBB would have experienced in the absence of the KlimaTicket. Comparing the outcomes of the actual and the synthetic ÖBB, I estimate that the public transport demand in Austria grew on average 6.3 or 3.3 percentage points (depending on the method) faster after the COVID-19 pandemic, mainly driven by higher growth rates in 2021. However, I also show that statistical significance is questionable, inter alia by replacing the outcome variable with Eurostat data on rail passengers from many European countries. To sum up, the causal effect of the KlimaTicket on public transport is positive but small and, from a statistical point of view, has to be taken with a grain of salt. 

My paper is structured as follows. Section \ref{Methods} outlines the study design and the methods used for the analysis. Moreover, Section \ref{Methods} also relates to institutional background information. Section \ref{Data} introduces the data and provides descriptive statistics. Section \ref{Results} presents the causal effect. Finally, Section \ref{Discussion and Conclusion} discusses and concludes.

\section{Identification and estimation}\label{Methods}

In my study, I aim to assess the treatment effect on a single treated unit (i.e., Austria) based on a comparison with multiple non-treated units. While I can directly observe the demand of the passengers in Austria, the identification of the Klimaticket's effect requires inferring the counterfactual outcome of how the public transport passengers in Austria would have evolved in the absence of the Klimaticket \citep[see, e.g., ][]{huber2023causal}. Therefore, I use the synthetic control method \citep[see, e.g., ][]{abadie2010synthetic,abadie2021using} to impute the counterfactual outcome under nontreatment based on an appropriate combination of other units (i.e., public transport companies) that are sufficiently similar to the ÖBB prior to the introduction, but did not experience any price reduction of a similar kind. The ÖBB's potential outcome under nontreatment is synthetically estimated as a weighted average of the observed post-treatment outcomes coming from the other units (the so-called donor pool), in my case other public transport companies from Europe. The non-treated units receive weights depending on how similar the public transport company was to the ÖBB prior to the price reduction. 

More formally, I analyze a panel data set of $n$ units. A unit's index is $i \in \{ 1,...,n \} $, which I observe over $T$ periods of time with a time index $t \in \{1,..,T\}$. $Y_{it}$ denotes unit's $i$ observed outcome in period $t$. In this study, only the last unit $i=n$ (i.e., the ÖBB) is treated in period $T_0+1$. Following \cite{huber2023causal}, I estimate a treatment effect $\hat{\Delta}_{n,T=t}$ for the treated unit $i=n$ in the post-treatment period $t \geq T_0+1$ (i.e., 2021 and 2022) as the difference of the treated outcome and a weighted average of non-treated units (i.e., comparable European railway companies) in that period. Given a set of weights, $W=(\hat{w}_1,...,\hat{w}_{n-1})$, I obtain the synthetic control estimator $\hat{\Delta}_{n,T=t}$: 

\begin{equation}
	\hat{\Delta}_{n,T=t}= Y_{nt}-\sum_{i=1}^{n-1} \hat{w}_iY_{it}, \text{for any } t \geq T_0+1. 
\end{equation}

Put simply, the idea of the synthetic control method is to choose the weights in a data-driven way such that the weighed average of pre-treatment outcomes of the non-treated observations mimics the evolution of the pre-treatment outcomes of the treated unit (i.e., the ÖBB). Applying the synthetic control method, I assume that I can appropriately model the post-treatment potential outcome under no treatment (i.e., no introduction of the KlimaTicket) by means of the synthetic ÖBB. Therefore, the weights of the non-treated units $i$ are choosen such that: 

\begin{equation}
	\sum_{i=1}^{n-1} \hat{w}_iY_{it} \approx Y_{nt}, \text{for all } t=1,....,T_0.
\end{equation}

Moreover, note that all weights together sum up to one, $\sum_{i=1}^{n-1} \hat{w}_i=1$, where the set of weights $W=(\hat{w}_1,...,\hat{w}_{n-1})$ is for any untreated unit $i$ is either positive or zero (i.e., $\hat{w}_i \geq 0$). 

Ensuring that my calculation identifies the effect of the KlimaTicket also relies on assumptions about the data-generating process, i.e., about how the world of public transport works (see, e.g., \cite{huntington2021effect}). Subsequently, I---following \cite{abadie2021using} and \cite{wallimann2023price}, the latter being the first study assessing the effect of a public transport price policy with the help of the synthetic control method---discuss the identifying assumptions underlying my analysis: \newline
\textbf{Assumption 1 (no anticipation):} \newline
According to Assumption 1, the demand in Austria must not change due to forward-looking passengers changing travel behavior prior to the KlimaTicket. To fulfill Assumption 1, I also consider 2021, the year of the introduction of the KlimaTicket, as a post-treatment period. That makes further sense, as federal states introduced regional offers on January 1, 2021. \newline
\textbf{Assumption 2 (availability of a comparison group):} \newline
Assumption 2 is satisfied when a so-called donor pool is sufficiently similar to the treated unit.\newline
\textbf{Assumption 3 (convex hull condition):} \newline
Assumption 3 implies that the pre-treatment outcomes of the treated unit are not too extreme compared to the non-treated units. That means the treatment unit's outcomes must not be (much) higher than the highest and lower than the lowest of the non-treated units during the pre-treatment period. \newline
\textbf{Assumption 4 (no spillover effects):} \newline
By Assumption 4, there must not be spillover effects from the KlimaTicket on other European transport companies. This seems reasonable, as people possessing the KlimaTIcket can (only) travel in the nationwide network of Austria. \newline
\textbf{Assumption 5 (no external shocks):} \newline
Assumption 5 is fulfilled when no external regional shocks occur to the outcome of interest (in one or only several treatment or control units) during the study period. In this study, I, for instance, assume that the COVID-19 pandemic and the following containment measures---as, for instance, quarantining having a deterrent influence on travel intentions \citep[see, e.g., ][]{husser2023comparative}---have comparable impacts on all companies (in different countries). 

The thing to notice is that I use the synthetic control method along with the synthetic difference-in-differences approach suggested by \citet{arkhangelsky2019synthetic}, which relies on the assumption that a weighted average of the outcomes of the control group in the pre-treatment period allows replicating the trend of the average potential outcome of the ÖBB under no treatment, i.e., no KlimaTicket. Put simply, the synthetic difference in difference approach arrives in a data-driven way, whether the conventional difference-in-differences or the synthetic control approach is more appropriate to approximate the outcome of the ÖBB in the absence of the KlimaTicket. 

To apply the synthetic control method in the statistical software \textsf{R}, I use the procedures of the \textit{synth} and \textit{SCtools} packages provided by \cite{hainmueller2015synthPackage} and \cite{PackageSCtools}. Moreover, I use the \textit{synthdid} package by \cite{Arkhagelsky2021synthdid} to apply the synthetic difference-in-differences method.

\section{Data}\label{Data}

The investigation of the Klimaticket relies on annual reports of European railway companies. With the help of a student project, I systematically searched for (and contacted companies to gather) relevant performance indicators for the ÖBB and the control group (the so-called donor pool; see also Assumption 2). It turns out that the number of passengers is the most feasible indicator for public transport demand in order to create a panel data set. However, for many companies, the relevant demand indicator was not available (or the email has been ignored), see Table \ref{table:contact} in Appendix \ref{Appendix_A}. 

\begin{figure}[H]
	\centering \caption{\label{Passanger_Dev} Passenger development since 2015}	\includegraphics[scale=.25]{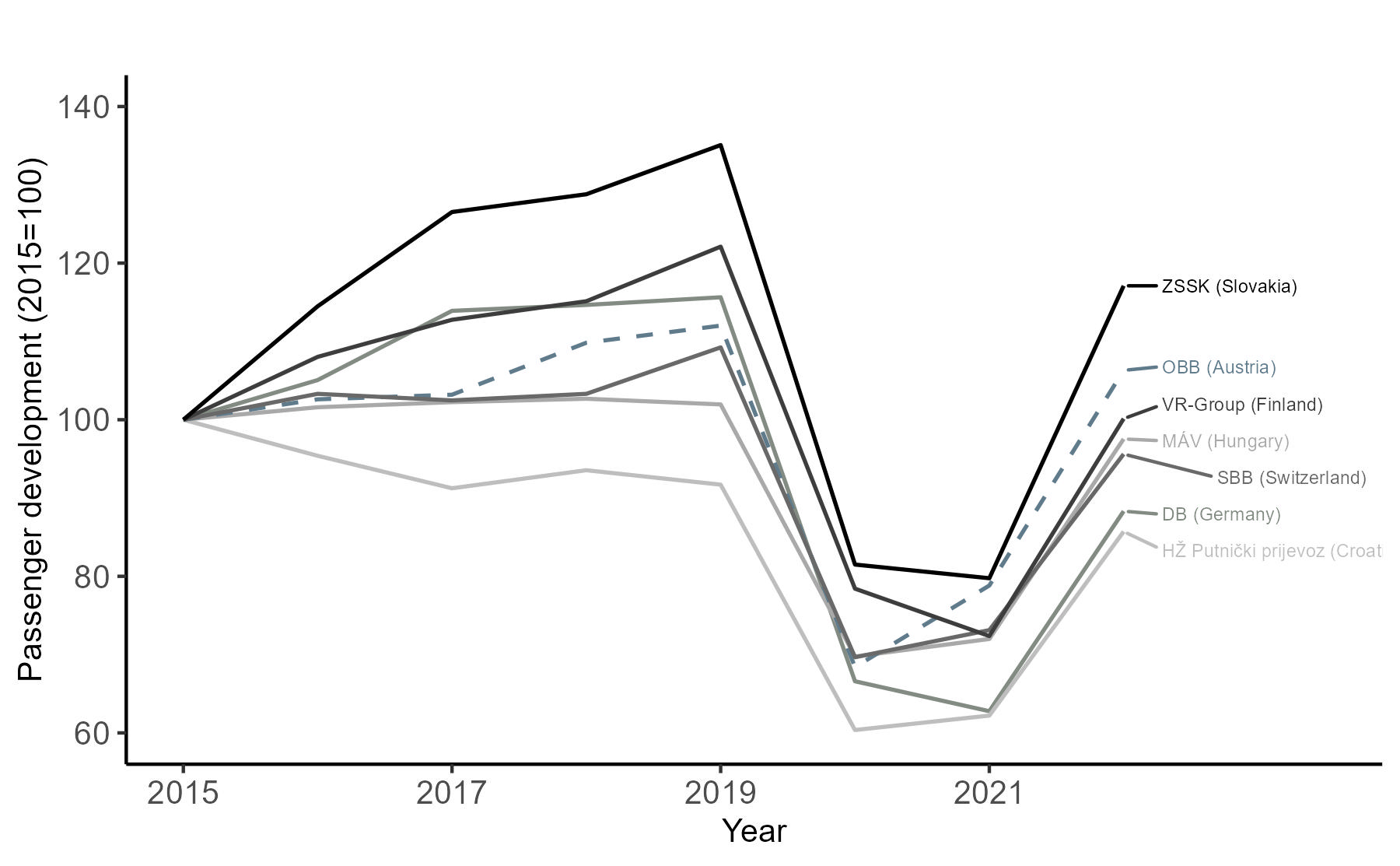}
\end{figure}

I obtain a data set with the ÖBB and six companies in the control group, i.e., DB (Germany), HŽ Putnički prijevoz (Croatia), MÁV (Hungary), SBB (Switzerland), VR-Group (Finland), and ZSSK (Slovakia). On average, from 2015 to 2022, the DB had 2158 million passengers, by far the highest number of passengers, followed by the SBB with 418 million and the ÖBB with 232 million. On the other hand, the HŽ Putnički prijevoz had, on average, 18.4 million, the lowest number of passengers. The ÖBB had an average of 232 million passengers. Since the difference in the outcome variables is enormous, I transform the outcome to growth rates, i.e., $(Y_{it}-Y_{it-1})/Y_{it-1}*100$ \citep[see, e.g., ][]{abadie2021using}. Figure \ref{Passanger_Dev} depicts the passenger developments of the previously mentioned companies from 2015 to 2022. 

Figure \ref{Passanger_Dev} indicates that the demand of all companies in the control and treatment group decreased during the COVID-19 pandemic and the following containment measures. Therefore, I assume the COVID-19 pandemic does not violate Assumption 5 (no external shock). However, to be more precise, I have to supplement the research question in the sense of: \textit{Has the KlimaTicket led to a better recovery of the ÖBB from the drop in demand caused by the COVID-19 pandemic?} When looking at the number of passengers shown in Figure \ref{Passanger_Dev}, I see that the numbers of all railway companies are recovering, and the recovery of the ÖBB (dashed line) is slightly above average. Moreover, since the 9-Euro-Ticket was available in Germany in 2022, this could impact the demand for the DB (even if not apparent in Figure \ref{Passanger_Dev}). Therefore, to account for Assumption 5, I ignore the DB data for the following analysis. 

As predictor variables to predict our outcome variable, I use pre-treatment outcomes for treated and control units from 2015 to 2020---we can think of pre-treatment outcomes serving as covariates to be controlled for. Moreover, I use aggregate Eurostat data on GDP and population growth to account for the potential of national public transport demand.\footnote{See, e.g., \hyperlink{https://ec.europa.eu/eurostat/databrowser}{https://ec.europa.eu/eurostat/databrowser}, accessed on December 25, 2023.}

Finally, in order to challenge the data source, I re-estimate the effect by replacing the outcome variable with the rail passenger data provided by Eurostat.\footnote{See, e.g., \hyperlink{https://ec.europa.eu/eurostat/databrowser/view/rail\_pa\_total}, accessed on January 17, 2024.} The thing to notice is that I now observe 19 control units (i.e., all European countries with more than 10 million passengers per year from 2015 to 2022 and non-confidential data), including Bulgaria, Croatia, Czechia, Denmark, Finland, France, Germany, Greece, Ireland, Italy, Latvia, Luxembourg, Norway, Portugal, Romania, Slovakia, Spain, Sweden, and Switzerland.  

\section{Results}\label{Results}

\subsection{Main results}

\begin{figure}[!htp]
	\centering \caption{\label{SCM} Demand development of the ÖBB and the synthetic ÖBB}	\includegraphics[scale=.5]{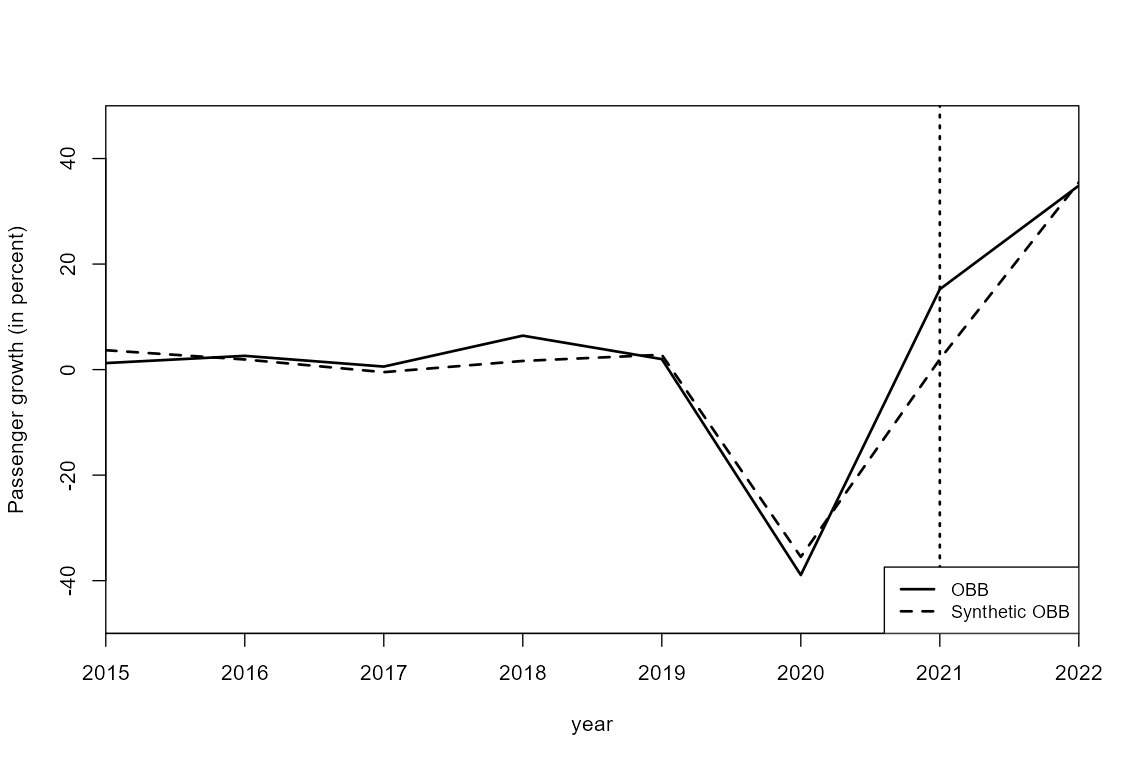}
\end{figure}

Figure \ref{SCM} summarises the main results when applying the synthetic control method. First, I see that the trajectories track each other closely during the pre-treatment period. Therefore, I conclude that the synthetic ÖBB is able to infer the counterfactual outcome of how the ÖBB would have evolved in the absence of the KlimaTicket. The synthetic control method chooses that all companies get nonzero weights: SBB with a weight of 40.7\%, HŽ Putnički prijevoz with 34.5\%, VR-Group with 14.2\%, ZSSK with 6.6\%, plus MÁV with 4.0\%. 

Second, Figure \ref{SCM} also shows the result of the KlimaTicket on the number of passengers. In 2021, I see that the demand for the ÖBB is above what it would have been without the KlimaTicket, i.e., 13.3 percentage points higher growth rate than without the KlimaTicket. However, considering that the KlimaTicket was launched on October 26, 2021, this difference could also be due to a faster COVID-19 pandemic recovery effect in Austria. While I observe a positive effect in 2021, the difference between the ÖBB and the synthetic ÖBB disappears in 2022. More precisely, the gap in 2022 amounts to -0.6 percentage points. Concluding, I observe no demand effect of the KlimaTicket in 2022 with the underlying study design and available data. On average, the effect amounts to higher growth (compared to the synthetic ÖBB) of 6.3 percentage points. The standard error amounts to 3.0 and the 95\% confidence interval to [0.5; 12.1] (both estimated with the \textit{synthdid} package). 

Applying the synthetic difference-in-differences method yields an average estimand of 3.3 percentage points higher demand growth (again on average across the two post-treatment periods 2021 and 2022) due to the KlimaTicket. However, the standard error amounts to 4.1 and is quite substantial relative to the absolute magnitude of the causal effect. Moreover, the 95\% confidence interval amounts to [-4.7; 11.4], concluding there might exist no relationship between the KlimaTicket and public transport demand. 

\subsection{Robustness investigations}

In order to investigate the robustness of my estimates, I substitute the growth rates with the actual number of passengers (see also Figure \ref{SCM_Einsteiger} in Appendix \ref{Appendix:Robustness}). When applying the synthetic control and synthetic difference-in-differences method, I see that the ÖBB had more passengers than it would have without the KlimaTicket, i.e., 29 and 27 million passengers more compared to the counterfactual, the synthetic ÖBB. The results again indicate a positive effect. However, the standard errors are even relatively higher than for the original result, with 88.6 and 521 million. 

As a second robustness check, I substitute the number of passengers from annual reports with the national rail passengers per country (see also Figure \ref{SCM_Eurostat} in Appendix \ref{Appendix:Robustness}). Interestingly, the effect of the KlimaTicket on demand growth amounts to -3.8 and 0.2 for the synthetic control and synthetic difference-in-differences method and is, therefore, even negative once. With again high standard errors of 14.4 and 11.4, when applying the Eurostat data, I conclude that we see no effect of the KlimaTicket on rail passengers. 

\section{Discussion and Conclusion}\label{Discussion and Conclusion}

In this study, I assessed the effect of the KlimaTicket on demand during the first and second years, right at the beginning and after the introduction. Therefore, I applied the synthetic control and synthetic difference-in-differences methods to mimic the ÖBBs' public transport demand growth rates in the absence of the KlimaTicket. The effect of the KlimaTicket on the COVID-19 pandemic recovery is, from an economic perspective, positive but small. Statistically speaking, considering the high standard errors (mainly when applying the synthetic difference-in-differences method), chance might drive the effect. Moreover, in 2022, the positive causal effect on demand growth disappears (and even becomes slightly negative). Finally, I have shown in a robustness investigation that there is no (positive) empirical effect of the KlimaTicket on passenger numbers if we look at the national level of railway travelers. 

However, as the theoretical mechanism of interest might have a long-term effect on demand, future studies should, with the help of the underlying study design of this case study, investigate the effect of the KlimaTicket in the upcoming years. These investigations will be valuable insofar as demand might increase over time \citep[see, for instance, ][]{wallimann2023price}. In the case of the KlimaTicket, this seems even more likely, as there is a high level of customer loyalty to the KlimaTicket, i.e., 80\% of customers repurchased the ticket after the first year \citep{KlimaTicketReport2022}. 

A limitation of this study is the data availability. Only data from five companies was available to create the synthetic counterfactual. However, by looking at the pre-treatment period in Figure \ref{SCM}, I conclude that I was able to create an accurate approximation of the ÖBB. In addition, regarding the data of the robustness check, I was also able to create a synthetic Austria that mimics the rail passenger growth in Austria. Moreover, I do not account (because of data availability reasons) for supply changes, e.g., frequency increases. Therefore, future studies should also account for supply changes in order to block off changing frequencies as an alternate explanation of the causal effect \citep[see, e.g., ][using an aggregate metric that inherits changes in public transport supply]{wallimann2023price}. 

Future research should also investigate whether the KlimaTicket encourages a transition from car-dependent individual transport systems to climate-friendly alternatives in light of the enormous effect of transport-related greenhouse gas emissions and (more context-related) as Austria aims to reduce private car usage by 16\% by 2024 \citep{OECD_Klimaticket_2022}. In a survey, 20\% of public transport users stated that they would have traveled by car without the KlimateTicket (5\% of the trips were induced) \citep{KlimaTicketReport2022}. In order to arrive at causal estimates, future studies could apply the underlying study design to assess the effect of the KlimaTicket on road traffic volumes. 

Finally, I want to add some thoughts to the design of the KlimaTicket, i.e., a season ticket with a single fixed fee regardless of usage---without discussing revenues of the ÖBB and following implications on public transport subsidies being a topic for future research. From a pricing psychological point of view, such an offer can make sense for several reasons \citep[see, e.g. ][]{wirtz2015flat,dang2024behavioural}. For instance, no so-called taximeter effect exists when a customer possesses the KlimaTicket, i.e., she does not have to travel with a pay-per-use tariff, making the costs of each journey transparent and thus less attractive. Moreover, purchasing the KlimaTicket makes budget planning for a customer easier. This links to the insurance effect, summarizing the finding that customers buy the KlimaTicket because they mentally value the agony of overspending more than the excitement of saving money (through pay-per-use traveling). 

On the other hand, a season ticket is---similar to a fare-free policy---no policy designed to reduce crowding during peak hours, making the utility of public transport travel less attractive \citep[see, e.g., ][]{batarce2016valuing,lu2024analyzing}. To adapt the public transport supply to peak demand (which can increase above average with a season ticket), infrastructure step fixed costs arise, i.e., costs that do not change within certain upper and lower thresholds but change (dramatically) when the thresholds are met. Considering the high railway infrastructure expenses---e.g., in 2021, the rail investment in Austria amounted to EUR 2,2 bn \citep{OECD_RailInvestiment}---measures reducing travel by public transport during peak hours such as fare price differentiation might be more advisable \citep[see, e.g., ][]{eriksson2023measures,huber2022business}.

	\newpage
	\bigskip
	
	\bibliographystyle{econometrica}
	\bibliography{Klimaticket.bib}
	
	\bigskip
	\newpage
	
\begin{appendix}
		
		\numberwithin{equation}{section}
		\counterwithin{figure}{section}
		\noindent \textbf{\LARGE Appendices}
	
	\section{Contacted companies}\label{Appendix_A}

	\begin{table}[H]
		\caption{European railway companies contacted (State December 2023)}\label{table:contact}
		\begin{center}
			\begin{tabular}{lcc}
				\hline
				\textbf{Country} & \textbf{Company}                                                     & \textbf{Number of passengers}                    \\\hline
				Austria          & ÖBB                                                                  & Yes                                    \\
				Belgium          & Société Nationale des Chemins de fer   Belges & No answer of railway company           \\
				Croatia          & Hrvatske   željeznice                                                & Yes                                    \\
				Czech   Republic & Ceskè dràhy                                                          & No                                     \\
				Denmark          &  Danske Statsbaner                           & No                                     \\
				Estonia          & Eesti Raudtee                                 & No answer of railway company           \\
				Finland          & VR Group                                                             & Yes                                     \\
				Germany 		 & DB & Yes \\
				Netherlands      & Nederlandse Spoorwegen                        & No answer of railway company           \\
				Hungary          & MÁV-START Co.                                                        & Yes                                    \\
				Italy            & FS Italiane                                                          & No                                     \\
				Latvia           & Latvijas dzelzceļš                            & No answer of railway company           \\
				Lithuania        & Lietuvos geležinkeliai                        & No answer of railway   company         \\
				Norway           & Vy Norges Statsbaner                          & No answer of railway   company         \\
				Poland           & Polskie Koleje   Państwowe                    & No answer of railway   company         \\
				Portugal         & Caminhos   de Ferro Portugueses, CP                                  & No                                     \\
				Serbia           & Srbija   Voz                                                         & No                                     \\
				Slovakia         & ZSSK									                                & Yes                                     \\
				Slovenia         & Slovenske   zeleznice                                                & No                                     \\
				Spain            & Renfe   Operadora                                                    & No                                     \\
				Sweden           & Statens   Järnvägar                                                  & No                                     \\
				Switzerland      & SBB                                          				        & Yes                                     \\
				United Kingdom   & Network   Rail                                                       & Yes, but only since 2020                            \\ \hline
			\end{tabular}
		\end{center}
	\end{table}

	\section{Robustness investigations}\label{Appendix:Robustness}
	
	\begin{figure}[!htp]
		\centering \caption{\label{SCM_Einsteiger} Passenger development of the ÖBB and the synthetic ÖBB }	\includegraphics[scale=.5]{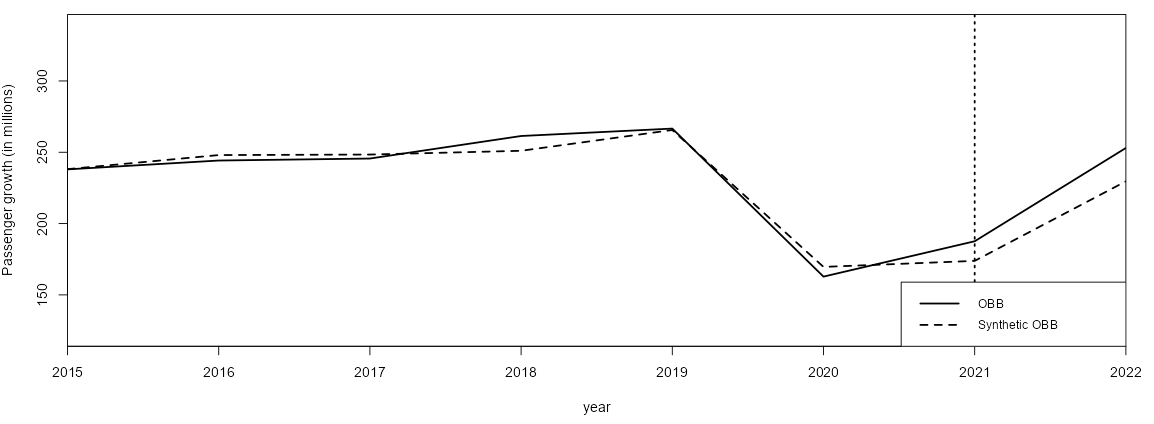}
	\end{figure}
	
	\begin{figure}[!htp]
		\centering \caption{\label{SCM_Eurostat} Rail demand development of Austria and the synthetic Austria (Eurostat data) }	\includegraphics[scale=.5]{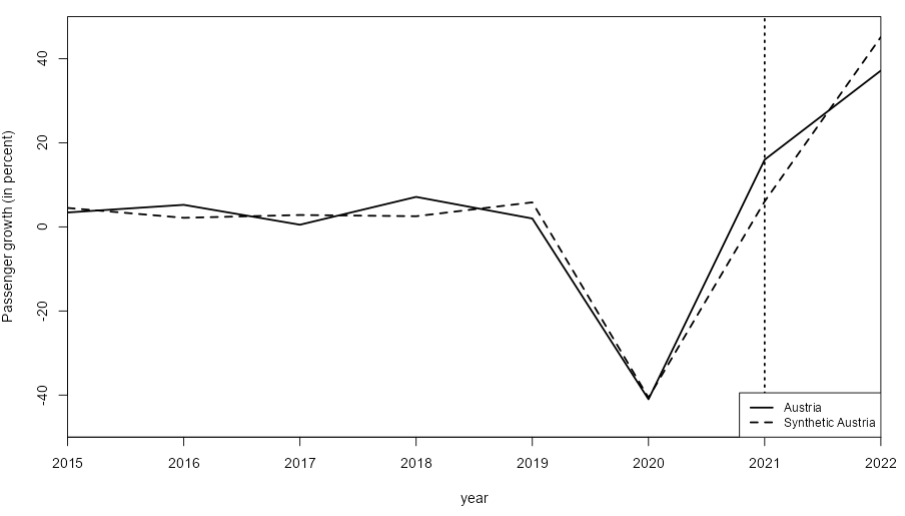}
	\end{figure}

	\end{appendix}
\end{document}